\documentclass[10pt, a4paper, twocolumn]{IEEEtran}\newcommand{\pictwidth}{80truemm}
\usepackage{amsmath,amssymb,verbatim,cite,amsopn,graphicx,citesort}

\DeclareMathOperator{\mvec}{vec}
\newcommand{\ve}[1]{\boldsymbol{#1}}
\newcommand{\E}[1]{E\left\{#1\right\}}
\newcommand{\vA}{\ve{A}}

\newcommand{\vF}{\ve{F}} 
 
\newcommand{\vH}{\ve{H}} \newcommand{\vh}{\ve{h}}
\newcommand{\vI}{\ve{I}} 
\newcommand{\vJ}{\ve{J}}

\newcommand{\vR}{\ve{R}}

 \newcommand{\vu}{\ve{u}}
 \newcommand{\vv}{\ve{v}}
\newcommand{\vW}{\ve{W}} \newcommand{\vw}{\ve{w}}
 \newcommand{\vx}{\ve{x}}
 \newcommand{\vy}{\ve{y}}

\newcommand{\powd}{\mathcal{P}}

\newcommand{\con}[1]{{#1}^{\ast}}

\newcommand{\mct}[1]{{#1}^{\dagger}}
\newcommand{\mt}[1]{{#1}^{T}}
\newcommand{\RxA}{n_R}
\newcommand{\TxA}{n_T}
\newcommand{\RxM}{M_R}
\newcommand{\TxM}{M_T}

\newcommand{\aod}{\phi}
\newcommand{\aoa}{\varphi}

\newcommand{\maod}{\aod_0}

\newcommand{\maoa}{\aoa_0}

\newcommand{\mcJ}{{\mathcal{J}}}
\newcommand{\snr}{\overline{\gamma}}

\newcommand{\RSISE}{Research School of Information Sciences and Engineering}
\newcommand{\ANU}{The Australian National University, ACT 0200, Australia.}
\newcommand{\AuthorOne}{Tharaka A. Lamahewa}
\newcommand{\AuthorTwo}{Rodney A. Kennedy}
\newcommand{\AuthorThree}{Thushara D. Abhayapala}
\newcommand{\AuthorFour}{Terence Betlehem}

\newcommand{\ThankOne}{This work was supported by the Australian Research Council
Discovery Grant DP0343804. }

\newcommand{\ThankTwo}{ R.A. Kennedy and T.D.
Abhayapala are also with National ICT Australia, Locked Bag 8001,
Canberra, ACT 2601, Australia. National ICT Australia is funded
through the Australian Government's \textit{Backing Australia's
Ability} initiative, in part through the Australian Research
Council.}

\newcommand{\conference}{Australian Communication Theory Workshop Proceedings 2006}
\title{MIMO Channel Correlation in General Scattering Environments}
\author{\authorblockN{\textit{\AuthorOne, \AuthorTwo, \AuthorThree\:and \AuthorFour}\\}
\authorblockA{
              \RSISE,\\
              \ANU}\\
\{tharaka.lamahewa, rodney.kennedy, thushara.abhayapala,
terence.betlehem\}@anu.edu.au.\thanks{\ThankOne}\thanks{\ThankTwo} }
\begin{document}
\maketitle
\begin{abstract}
This paper presents an analytical model for the fading channel
correlation in general scattering environments. In contrast to the
existing correlation models, our new approach treats the scattering
environment as non-separable and it is modeled using a bi-angular
power distribution. The bi-angular power distribution is
parameterized by the mean departure and arrival angles, angular
spreads of the univariate angular power distributions at the
transmitter and receiver apertures, and a third parameter, the
covariance between transmit and receive angles which captures the
statistical interdependency between angular power distributions at
the transmitter and receiver apertures. When this third parameter is
zero, this new model reduces to the well known ``Kronecker" model.
Using the proposed model, we show that Kronecker model is a good
approximation to the actual channel when the scattering channel
consists of a single scattering cluster. In the presence of multiple
remote scattering clusters we show that Kronecker model over
estimates the performance by artificially increasing the number of
multipaths in the channel.
\end{abstract}

%
%
\section{Introduction}\label{sec:introduction}
Wireless channel modelling has received much attention in recent
years since space-time processing using multiple antennas is
becoming one of the most promising areas for improvements in
performance of mobile communication systems
\cite{foschini-1998,tarokh_1998_first}. In channel modelling
research, the effects of fading channel correlation due to
insufficient antenna spacing and sparse scattering environments are
of primary concern as they impact the performance of multiple-input
multiple-output (MIMO) communication systems.

A popular channel model that has been used in MIMO performance
analysis is the ``Kronecker" model
\cite{da_shan_2000,gesbert_mimo,Kermoal-2002-MIMO}. In this model,
the correlation properties of the MIMO channel are modeled at the
transmitter and receiver separately, neglecting the statistical
interdependency between scattering distributions at the transmitter
and receiver antenna apertures. Measurement and analytical results
presented in \cite{bonek_kroneker_measurements,pollock_kronecker}
suggest that the Kronecker model does not accurately model the
underlying scattering channel, therefore it does not provide
accurate performance results.

In this paper, using a recently proposed spatial channel model
\cite{abhayapala_2003_channel}, we develop an alternate to the
Kronecker model which gives channel correlation for a general class
of scattering environments. In our proposed model, fading channel
correlation is parameterized by the antenna configuration details
(spacing and the placement) both at the transmitter and the receiver
arrays, and the joint bi-angular power distribution between
transmitters and receivers, which models the scattering environment
surrounding the transmit and receive antenna apertures. The
bi-angular power distribution is  parameterized by the mean
departure and arrival angles, angular spreads of the univariate
angular power distributions at the transmitter and receiver
apertures, and a third parameter, the covariance between transmit
and receive angles which captures the statistical interdependency
between angular power distributions at the transmitter and receiver
apertures. When this third parameter is zero, i.e., the power
distribution at the transmitter is independent of the power
distribution at the receiver, the proposed correlation model reduces
to the Kronecker model. In order to model the scattering environment
we propose several bi-angular power distributions and also find the
correlation coefficients associated with these distributions in
closed form. Using the proposed model, we show that Kronecker model
is a good approximation to the actual channel when the scattering
channel consists of a single scattering cluster. We also show that
when the scattering channel consists of multiple remote scattering
clusters, the Kronecker model over estimates the performance MIMO
systems by artificially increasing the number of scattering clusters
in the scattering channel.

\textbf{\\Notations:} Throughout the paper, the following notations
will be used: Bold lower (upper) letters denote vectors (matrices).
$\mt{[\cdot]}$, $\con{[\cdot]}$ and $\mct{[\cdot]}$ denote the
transpose, complex conjugate and conjugate transpose operations,
respectively. The symbols $\delta(\cdot)$ and $\otimes$ denote the
Dirac delta function and Matrix Kronecker product, respectively. The
notation $\E{\cdot}$ denotes the mathematical
expectation,$\mvec({\vA})$ denotes the vectorization operator which
stacks the columns of $\vA$, $\lceil{.}\rceil$ denotes the
ceiling operator and $\mathbb{S}^1$ denotes the unit circle.\\

\section{Spatial Channel Model}\label{sec:spatial_channel_model}
First we review the spatial channel model proposed in
\cite{abhayapala_2003_channel}. Consider a MIMO system consisting of
$\TxA$ transmit antennas located at positions $\vx_t$,
$t=1,2,\cdots,\TxA$ relative to the transmitter array origin, and
$\RxA$ receive antennas located at positions $\vy_r$,
$r=1,2,\cdots,\RxA$ relative to the receiver array origin. $r_T \geq
\max\parallel\!\!\vx_t\!\!\parallel$ and $r_R \geq
\max\parallel\!\!\vy_r\!\!\parallel$ denote the radius of spheres
that contain all the transmitter and receiver antennas,
respectively. We assume that scatterers are distributed in the far
field from the transmitter and receiver antennas and regions
containing the transmit and receive antennas are distinct.


By taking into account physical aspects of scattering, the MIMO
channel matrix $\vH$ can be decomposed into deterministic and random
parts as \cite{abhayapala_2003_channel}
\begin{align}\label{eqn:channel_decompo}
\vH &= \vJ_R\vH_S\mct{\vJ}_T,
\end{align}
where $\vJ_R$ is the receiver configuration matrix,
\begin{align}\nonumber
\vJ_R &= \left[
  \begin{array}{ccc}
    \mcJ_{-\RxM}(\vv_1) & \cdots & \mcJ_{\RxM}(\vv_1) \\
    \mcJ_{-\RxM}(\vv_2) & \cdots & \mcJ_{\RxM}(\vv_2) \\
    \vdots & \ddots & \vdots \\
    \mcJ_{-\RxM}(\vv_{\RxA}) & \cdots & \mcJ_{\RxM}(\vv_{\RxA}) \\
  \end{array}
\right],
\end{align}
and $\vJ_T$ is the transmitter configuration matrix,
\begin{align}\nonumber
\vJ_T &= \left[
  \begin{array}{ccc}
    \mcJ_{-\TxM}(\vu_1) & \cdots & \mcJ_{\TxM}(\vu_1) \\
    \mcJ_{-\TxM}(\vu_2) & \cdots & \mcJ_{\TxM}(\vu_2) \\
    \vdots & \ddots & \vdots \\
    \mcJ_{-\TxM}(\vu_{\TxA}) & \cdots & \mcJ_{\TxM}(\vu_{\TxA}) \\
  \end{array}
\right],
\end{align}
 where
$\mcJ_{n}(\vx)$ is the spatial-to-mode function (SMF) which maps the
antenna location to the $n$-th mode of the region. The form which
the SMF takes is related to the shape of the scatterer-free antenna
region. For a circular region in 2-dimensional space, the SMF is
given by a Bessel function of the first kind
\cite{abhayapala_2003_channel} and for a spherical region in
3-dimensional space, the SMF is given by a spherical Bessel function
\cite{abhayapala_2003_3Dchannel}. For a prism-shaped region, the SMF
is given by a prolate spheroidal function \cite{Hanlen_2003}.

Here we consider the situation where the multipath is restricted to
the azimuth plane only (2-D scattering environment), having no field
components arriving at significant elevations. In this case, the SMF
is given by
\begin{align}
\mcJ_{n}(\vw)\,&{\triangleq}\,J_n(k\|\vw\|)e^{{\imath}n(\aod_w-\pi/2)}\nonumber,
\end{align}
where $J_n(\cdot)$ is the Bessel function of integer order $n$,
vector $\vw\:\equiv\:(\|\vw\|,\aod_w)$ in polar coordinates is the
antenna location relative to the origin of the aperture which
encloses the antennas, $k=2\pi/\lambda$ is the wave number with
$\lambda$ being the wave length and $\imath=\sqrt{-1}$. $\vJ_T$ is
$\TxA{\times}(2\TxM+1)$ and $\vJ_R$ is $\RxA{\times}(2\RxM+1)$,
where $2\TxM+1$ and $2\RxM+1$ are the number of
effective\footnote{Although there are infinite number of modes
excited by an antenna array, there are only finite number of modes
$(2M+1)$ which have sufficient power to carry information.}
communication modes at the transmit and receive regions,
respectively. Note, $\TxM$ and $\RxM$ are defined by the size of the
regions containing all the transmit and receive antennas,
respectively \cite{H_Jones_2002_b}. In our case,
\begin{align}\nonumber
\TxM&=\left\lceil{\frac{ker_T}{\lambda}}\right\rceil\text{\quad and
}
\RxM=\left\lceil{\frac{ker_R}{\lambda}}\right\rceil,\nonumber
\end{align}
where $e\approx{2.7183}$.

Finally, $\vH_S$ is the $(2\RxM+1)\times(2\TxM+1)$ random complex
scattering channel matrix with $(\ell,m)$-th element given by
\begin{align}\label{eqn:H_S}
\left\{\vH_S\right\}_{\ell,m} &=
\iint_{\mathbb{S}^1\times\mathbb{S}^1}\!g(\aod,\aoa)e^{\imath(m-\TxM-1)\phi}
e^{-\imath(\ell-\RxM-1)\psi} \mathrm{d\aod}\mathrm{d\aoa}
\end{align}
representing the complex scattering gain between the $(m-\TxM-1)$-th
mode of the scatter-free transmit region and $(\ell-\RxM-1)$-th mode
of the scatter-free receiver region, where $g(\aod,\aoa)$ is the
effective random complex scattering gain function  for signals with
angle-of-departure $\aod$ from the scatter-free transmitter region
and angle-of-arrival $\aoa$ at the scatter-free receiver region. The
reader is referred to \cite{abhayapala_2003_channel} for more
information about the channel decomposition \eqref{eqn:channel_decompo}.\\


The correlation matrix of the MIMO channel $\vH$ given by
\eqref{eqn:channel_decompo} can be written as
\begin{align}\nonumber
\vR=\E{\vh\mct{\vh}} =
\left(\con{\vJ}_T\otimes\vJ_R\right)\vR_S(\mt{\vJ}_T\otimes\mct{\vJ}_R),
\end{align}
where $\vh=\mvec{(\vH)}$ and $\vR_S$ the modal correlation matrix of
the scattering channel,
\begin{align}\nonumber
\vR_S=\E{\vh_S\mct{\vh}_S},
\end{align}
with $\vh_S=\mvec{(\vH_S)}$. Modal correlation matrix $\vR_S$ can
also be written as a block matrix of $(2\TxM+1)\times(2\TxM+1)$
blocks, each of size $(2\RxM+1)\times(2\RxM+1)$,
\begin{align}\nonumber
\vR_S &\!=\! \left[
  \begin{array}{cccc}
    \!\!\vR_{S,1,1}\!       & \vR_{S,1,2}\!       &\!\cdots\! & \vR_{S,1,2\TxM+1}\!\!       \\
    \!\!\vR_{S,2,1}\!       & \vR_{S,2,2}\!       &\!\cdots\! & \vR_{S,2,2\TxM+1}\!\!       \\
    \!\!\vdots\!            & \vdots\!            &\!\ddots\! & \vdots\!\!                  \\
    \!\!\vR_{S,2\RxM+1,1}\! & \vR_{S,2\RxM+1,2}\! &\!\cdots\! & \vR_{S,2\RxM+1,2\TxM+1}\!\! \\
  \end{array}
\right],
\end{align}
where $\vR_{S,m,m'}$ is the correlation between $m$-th and $m'$-th
columns of $\vH_S$. A diagonal block $\vR_{S,m,m}$ gives the modal
correlation matrix at the receiver region due to the $m$-th mode at
the transmit region whereas off diagonal blocks $\vR_{S,m,m'}$ give
the cross correlation between two distinct modal pairs at the
transmit and receiver apertures.

\subsection{Modal Correlation in General Scattering
Environments}\label{sec:modal_correlation_matrix} Using
\eqref{eqn:H_S}, we can define the modal correlation between complex
scattering gains as
\begin{align}\label{eqn:modal_corr_def}
\gamma_{m,m'}^{\ell,\ell'}&\triangleq
\E{\{\vH_S\}_{\ell,m}\con{\{\vH_S\}}_{\ell',m'}}.
\end{align}
Substituting \eqref{eqn:H_S} in \eqref{eqn:modal_corr_def} gives
\begin{align}\label{eqn:joint_modal_correlation}%
\gamma_{m,m'}^{\ell,\ell'} \!\!&=\!\!
\int_{4}\!\E{g(\aod,\aoa)g^*(\aod',\aoa')}\!e^{i(m-\TxM-1)\aod}e^{-i(m'-\TxM-1)\aod'}\nonumber\\
&e^{-i(\ell-\RxM-1)\aoa}e^{i(\ell'-\RxM-1)\aoa'}
\mathrm{d\aod}\mathrm{d\aoa}\mathrm{d\aod'}\mathrm{d\aoa'},
\end{align}
where we have introduced the shorthand
$\int_{4}{\triangleq}\int\!\!\int\!\!\int\!\!\int_{\mathbb{S}^1\times\mathbb{S}^1}$.

Assume that the scattering from one direction is independent of that
from another direction for both the receiver and the transmitter
apertures (WSSUS). Then the second-order statistics of the
scattering gain function $g(\aod,\aoa)$ is given by
\begin{align}\nonumber
\E{g(\aod,\aoa)\con{g}({\aod}^{'},{\aoa}^{'})} &=
G(\aod,\aoa)\delta(\aod-{\aod}^{'})\delta(\aoa-{\aoa}^{'}),
\end{align}
where $G(\aod,\aoa)=\E{|g(\aod,\aoa)|^2}$ is the 2D \textit{joint
azimuth power spectral density\footnote{also called bi-angular power
distribution or joint scattering distribution.}} (PSD) over
departure and arrival angles $\aod$ and $\aoa$ of the scattering
channel, normalized such that the total scattering channel power
\begin{align}\nonumber
\iint_{\mathbb{S}^1\times\mathbb{S}^1}G(\aod,\aoa)d\aoa{d\aod} = 1.
\end{align}
Using this assumption, the modal correlation coefficient
\eqref{eqn:joint_modal_correlation} can then be simplified to
\begin{align}\label{eqn:modal_corr_gen}
\gamma_{m,m'}^{\ell,\ell'}&=\gamma_{m-m'}^{\ell-\ell'},\nonumber\\
&=\iint_{\mathbb{S}^1\times\mathbb{S}^1}G(\aod,\aoa)e^{-i(\ell-\ell')\aoa}e^{i(m-m')
\aod}d\aoa{d\aod},
\end{align}
which gives the $(\ell,\ell')$-th element
of $\vR_{S,m,m'}$.\\

Since the scattering gain function $g(\aod,\aoa)$ is periodic in
both $\aod$ and $\aoa$, the joint PSD function $G(\aod,\aoa)$ is
also periodic in both $\aod$ and $\aoa$. Therefore, using the
orthogonal circular harmonics $e^{in\aoa}$ as the basis
set, 
 $G(\aod,\aoa)$ can be expanded in a 2-D Fourier series as,
\begin{align}\nonumber
G(\aod,\aoa) &=
\frac{1}{4\pi^2}\sum_{p=-\infty}^{\infty}\sum_{q=-\infty}^{\infty}{\beta_q^p}
e^{-iq\aod}e^{ip\aoa},
\end{align}
with coefficients
\begin{align}\label{eqn:Fourier_series_coeff}
\beta_q^p &= \iint_{\mathbb{S}^1\times\mathbb{S}^1}G(\aod,\aoa)
e^{-ip\aoa}e^{iq\aod}\mathrm{{d\aoa}}\mathrm{{d\aod}}.
\end{align}

By comparing \eqref{eqn:Fourier_series_coeff} with
\eqref{eqn:modal_corr_gen}, we can see that entries of $\vR_S$ are
given by 2-D Fourier coefficients of the joint PSD function
$G(\aod,\aoa)$. Furthermore, from \eqref{eqn:modal_corr_gen}, it is
evident that the entries of $\vR_S$ are dependent on the joint PSD
$G(\aod,\aoa)$, which is usually parameterized by the mean departure
and arrival angles $\maod$, $\maoa$, angular spreads $\sigma_t$ and
$\sigma_r$ for distributions at the transmitter and receiver
apertures, and the covariance
\begin{align}\label{eqn:tx_rx_angles_cov}
\rho=\mathrm{cov}(\aod,\aoa)\triangleq\frac{\E{\aod\aoa}-\maod\maoa}{\sigma_t\sigma_r}
\end{align}
between transmit and receive angles.

From \eqref{eqn:modal_corr_gen}, the correlation between $\ell$-th
and $\ell^{\prime}$-th modes at the receiver region due to the
$m$-th mode at the transmitter region is given by
\begin{align}\label{eqn:rx_modal_corr_coef}
\gamma^{\ell,\ell^{\prime}}&={\int_{\mathbb{S}^1}}{\powd}_{Rx}(\aoa)e^{-i(\ell-\ell^{\prime})\aoa}d\aoa,
\end{align}
where ${\powd}_{Rx}(\aoa)$ is the average power density of the
scatterers surrounding the receiver region, given by the
marginalized PSD
\begin{align}\nonumber
{\powd}_{Rx}(\aoa)={\int_{\mathbb{S}^1}}G(\aod,\aoa){d\aod}.
\end{align}
Similarly, the correlation between $m$-th and $m^{\prime}\!$-th
modes at the transmitter region due to the $\ell$-th mode at the
receiver region is given by
\begin{align}\label{eqn:tx_modal_corr_coef}
\gamma_{m,m^{\prime}}&={\int_{\mathbb{S}^1}}{\powd}_{Tx}(\aod)e^{i(m-m^{\prime})\aod}d\aod,
\end{align}
where ${\powd}_{Tx}(\aod)={\int_{\mathbb{S}^1}}G(\aod,\aoa){d\aoa}$
is the average power density of the scatterers surrounding the
transmitter region.

\subsection{Kronecker Model as a Special
Case}\label{sec:kronecker_special_case} When the covariance between
departure and arrival angles is zero, $\rho=0$, the joint PSD can
then be expressed as the product of scattering distributions at the
transmitter and receiver regions, i.e.,
\begin{align}\nonumber
G(\aod,\aoa) &={\powd}_{Tx}(\aod){\powd}_{Rx}(\aoa).
\end{align}

This separability condition leads to the well known `Kronecker'
model\cite{Kermoal-2002-MIMO,pollock_kronecker,weichselberger_channel},
where we have
\begin{align}\nonumber
\vR_S = \vF_T\otimes\vF_R
\end{align}
with $\vF_T$ the $(2\TxM+1)\times(2\TxM+1)$ transmitter modal
correlation matrix and $\vF_R$ the $(2\RxM+1)\times(2\RxM+1)$
receiver modal correlation matrix. The $(m,m')$-th element of
$\vF_T$ is given by \eqref{eqn:tx_modal_corr_coef} and the
$(\ell,\ell')$-th element of $\vF_R$ is given by
\eqref{eqn:rx_modal_corr_coef}.

The separability of $G(\aod,\aoa)$ when $\rho=0$ also yields that
\begin{itemize}
  \item modal correlation at the transmitter $\gamma_{m,m'}$ is
  independent of the mode selected from the receiver region,
  \item modal correlation at the receiver $\gamma^{\ell,\ell'}$ is independent of the mode
  selected from the transmitter region and
  \item correlation between two distinct modal pairs is the
  product of corresponding modal correlations at the transmitter and \
  the receiver,
  \begin{align}\nonumber
  \gamma_{m,m'}^{\ell,\ell'}&=\gamma_{m,m'}\gamma^{\ell,\ell'}.
  \end{align}
\end{itemize}

In the Kronecker model, diagonal blocks $\vR_{S,m,m}$ and
off-diagonal blocks $\vR_{S,m,m'}$ of $\vR_S$ are given by $\vF_R$
and $\gamma_{m,m'}\vF_R$, respectively.

\section{Bi-Angular Scattering
Distributions}\label{sec:bidirectional_power_dists} In this section,
we outline several examples of bi-angular scattering distributions
along with their modal correlation coefficients
\eqref{eqn:modal_corr_gen}, which give the entries of the modal
correlation matrix $\vR_S$.

\subsection{Uniform Limited azimuth
field}\label{sec:uniform_limited} When the energy leaves uniformly
to a restricted range of azimuth
$(\maod-\bigtriangleup_t,\maod+\bigtriangleup_t)$ at the transmitter
 and to arrive at the receiver uniformly from $(\maoa-\bigtriangleup_r,\maoa+\bigtriangleup_r)$, then
following Morgenstern's family of distributions
\cite{morgenstern_family_dist}, we have the joint uniform limited
azimuth scattering distribution
\begin{align}\nonumber
G_{U}(\aod,\aoa) &=
    \frac{1}{4\bigtriangleup_t\bigtriangleup_r} - \frac{\rho(\aod-\maod)(\aoa-\maoa)}
    {4\bigtriangleup_t^2\bigtriangleup_r^2},
\end{align}
for $|\aod-\maod|\leq{\bigtriangleup_t}$ and
$|\aoa-\maoa|\leq{\bigtriangleup_r}$, and $0$ elsewhere. The
parameter $\rho$ is the covariance between $\aod\in[-\pi,\pi)$ and
$\aoa\in[-\pi,\pi)$, which controls the flatness of
$G_{U}(\aod,\aoa)$. In this case, the modal correlation coefficients
\eqref{eqn:modal_corr_gen} are given by,
\begin{align}\nonumber
\gamma_{m,m'}^{\ell,\ell'}\!\!&=\! \left\{
  \begin{array}{ll}
    \!\!\mathrm{sinc}((m-m')\Delta_t)e^{i(m-m')\maod}, & \!\!\!\!\hbox{if $\ell\!=\!\ell'$
    and $m\!\neq\!{m'}$} \\
    \!\!\mathrm{sinc}((\ell-\ell')\Delta_r)e^{-i(\ell-\ell')\maoa}, & \!\!\!\!\hbox{
    if $\ell\!\neq\!{\ell'}$ and $m\!=\!m'$} \\
    \!\!e^{i((m-m')\maod-(\ell-\ell')\maoa)}{\Gamma}_{m,m'}^{\ell,\ell'}, & \!\!\!\!\hbox{otherwise}
  \end{array}
\right.
\end{align}
where ${\Gamma}_{m,m'}^{\ell,\ell'}$ is given by
\eqref{eqn:uniform_limted_gamma}. Note that $G_U(\aod,\aoa)$ has
marginal distributions $\powd_{Tx}(\aod)=1/2\bigtriangleup_t$ for
$\aod\in(\maod-\bigtriangleup_t,\maod+\bigtriangleup_t)$ and zero
elsewhere, and $\powd_{Rx}(\aoa)=1/2\bigtriangleup_r$ for
$\aoa\in(\maoa-\bigtriangleup_r,\maoa+\bigtriangleup_r)$ and zero
elsewhere, with corresponding transmit and receive modal correlation
coefficients
$\gamma_{m,m'}=\mathrm{sinc}((m-m')\Delta_t)e^{i(m-m')\maod}$ and
$\gamma^{\ell,\ell'}=\mathrm{sinc}((\ell-\ell')\Delta_r)e^{-i(\ell-\ell')\maoa}$,
respectively. For $\bigtriangleup_t=\pi$ and $\bigtriangleup_r=\pi$
with $\rho=0$ (isotropic scattering), we have uniform PSD
$G_U(\aod,\aoa)=1/4\pi^2$ and the modal correlation coefficients
become,
\begin{align}\nonumber
\gamma_{m,m'}^{\ell,\ell'} = \delta_{m-m'}\delta_{\ell-\ell'},
\end{align}
corresponding to i.i.d. elements of $\vH_S$. In this special case,
the correlation matrix $\vR$ takes the form
\begin{align}\nonumber
\vR= (\con{\vJ}_T\mt{\vJ}_T)\otimes(\vJ_R\mct{\vJ}_R).
\end{align}

\begin{figure*}[!t]
\normalsize
\begin{align}\label{eqn:uniform_limted_gamma}
 {\Gamma}_{m,m'}^{\ell,\ell'} &= \mathrm{sinc}((m-m')\Delta_t)\mathrm{sinc}((\ell-\ell')\Delta_r)
+\frac{{\rho}}
{(\ell-\ell')(m-m')\Delta_t\Delta_r}\times\nonumber\\
&\left[\cos((m-m')\Delta_t)-\mathrm{sinc}((m-m')\Delta_t)\right]
\left[\mathrm{sinc}((\ell-\ell')\Delta_r)-\cos((\ell-\ell')\Delta_r)\right].
\end{align}
\hrulefill
\end{figure*}

\subsection{Truncated Gaussian Distributed
Field}\label{sec:truncated_gaussian}
A distribution that can be used to model the joint PSD is the
truncated Gaussian bivariate distribution, defined as
\begin{align}\nonumber
G_G(\aod,\aoa) &=\Omega_G\exp\left[
\frac{-Q(\aod,\aoa)}{2(1-\rho^2)}\right], \qquad\aod, \aoa
\in[-\pi,\pi),
\end{align}
where $\Omega_G$ is a normalization constant such that
$\int\int_{\mathbb{S}^1}G_G(\aod,\aoa)d\aod{d\aoa}=1$ and
\begin{align}\nonumber
Q(\aod,\aoa)\! = \!\frac{(\aod\!-\!\maod)^2}{\sigma_t^2}-
\frac{2\rho(\aod\!-\!\maod)(\aoa\!-\!\maoa)}{\sigma_t\sigma_r}+
\frac{(\aoa\!-\!\maoa)^2}{\sigma_r^2}
\end{align}
with $\maod$ the mean AOD at the transmitter, $\sigma_t$ the
standard deviation of the non-truncated marginalized PSD at the
transmitter, $\maoa$ the mean AOA at the receiver, $\sigma_r$ the
standard deviation of the non-truncated marginalized PSD at the
receiver and $\rho$ is the covariance between receive and transmit
angles, as defined by \eqref{eqn:tx_rx_angles_cov}. In this case,
finding modal correlation coefficients in closed form poses a much
harder problem. However, if the angular spread at the both end of
the channel is small, then a good approximation for the truncated
Gaussian case can be obtained by integrating over the domain
$(-\infty,\infty)$, since the tails of marginalized PSDs cause a
very little error. Using a result found in \cite{lukacs_book_70},
\begin{small}
\begin{align}\nonumber
\gamma_{m,m'}^{\ell,\ell'}{\approx}&
\exp(i((m\!-\!m')\maod-(\ell\!-\!\ell')\maoa)-\nonumber\\
&\frac{1}{2}\left(\sigma_t^2(m\!-\!m')^2-
2\rho\sigma_t\sigma_r(m\!-\!m')(\ell\!-\!\ell')+\sigma_r^2(\ell\!-\!\ell')^2\right))\nonumber.
\end{align}
\end{small}

\vspace{-0.7cm}
\subsection{Truncated Laplacian Distributed
Field}\label{sec:truncated_laplace}
Similar to the truncated Gaussian distribution, an elliptical
truncated bivariate Laplacian distribution can be defined as
\cite{Kozubowski_bi_laplacian}
\begin{align}\nonumber
G_L(\aod,\aoa) &={\Omega_L}K_0\left(
\sqrt{\frac{2Q(\aod,\aoa)}{(1-\rho^2)}}\right), \qquad\aod, \aoa
\in[-\pi,\pi),
\end{align}
where $\Omega_L$ is a normalization constant such that
$\int\int_{\mathbb{S}^1}G_G(\aod,\aoa)d\aod{d\aoa}=1$ and
$K_0(\cdot)$ is the modified Bessel function of the second kind of
order zero. The modal correlation coefficients
\eqref{eqn:modal_corr_gen} for this distribution are given by
\begin{small}
\begin{align}\nonumber
\gamma_{m,m'}^{\ell,\ell'}{=}&
\frac{\exp(i(m\!-\!m')\maod-(\ell\!-\!\ell')\maoa)}{\sigma_t^2(m\!-\!m')^2-
2\rho\sigma_t\sigma_r(m\!-\!m')(\ell\!-\!\ell')+\sigma_r^2(\ell\!-\!\ell')^2+1}.
\end{align}
\end{small}

\section{Simulation Examples}\label{sec:simulation_examples}
In this section we compare the performance of MIMO communication
systems operating in separable (kronecker channel with $\rho=0$ in
\eqref{eqn:modal_corr_gen}) and non-separable scattering
environments.

We consider transmit and receive apertures of radius $0.5\lambda$,
corresponding to $2\lceil{{\pi}e0.5}\rceil+1=11$ effective
communication modes at each aperture. Within each aperture, we place
three antennas in a uniform circular array (UCA) configuration
($3\times{3}$ MIMO channel). The system performance is measured in
terms of the average mutual information. Here we assume transmitter
has no knowledge about the channel and the receiver has the full
knowledge about the channel. In this case, the average mutual
information is given by
\begin{align}\nonumber
\mathcal{I}
&=\E{\log_2\left|\vI_{\RxA}+\frac{\snr}{\TxA}\vH\mct{\vH}\right|},
\end{align}
where $\snr$ is the average symbol energy-to-noise ratio (SNR) at
each receiver antenna.

Assuming $\vR$ is a positive definite matrix, a realization of the
MIMO channel $\vH$ is obtained by forming
\begin{align}\nonumber
\mvec{(\vH)} &= \vR^{1/2}\mvec{(\vW)},
\end{align}
where $\vR^{1/2}$ is the positive definite matrix square root of
$\vR$ and $\vW$ is a $\RxA\times\TxA$ matrix which has zero-mean
independent and identically distributed complex Gaussian random
entries with unit variance.

\begin{figure}[h]
\centering
\includegraphics[width=\pictwidth]{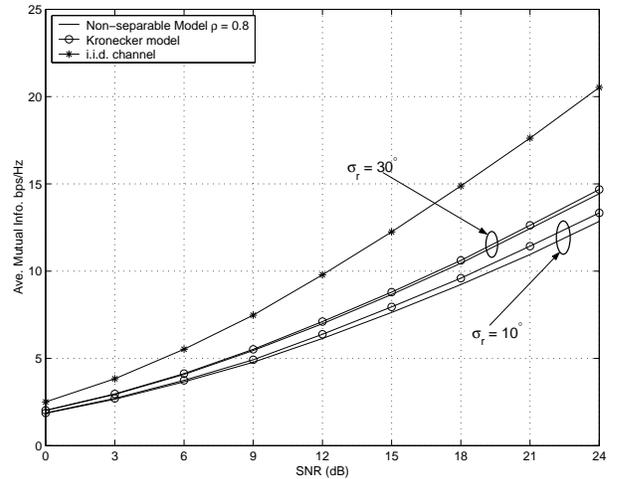}
\caption{Average mutual information of 3-transmit UCA and 3-receive
UCA MIMO system in separable (Kronecker with $\rho=0$) and
non-separable $(\rho=0.8)$ scattering environments: bivariate
truncated Gaussian azimuth field with mean AOD $=90^\circ$, mean AOA
$=90^\circ$, transmitter angular spread $\sigma_t=10^\circ$ and
receiver angular spreads
$\sigma_r=\{30^\circ,10^\circ\}$.}\vspace{-0.25cm}
\label{fig:unimodal_gauss_cap_sigr_30_10}
\end{figure}


Figure \ref{fig:unimodal_gauss_cap_sigr_30_10} shows the average
mutual information for a bivariate truncated Gaussian distributed
azimuth field with $\rho=0.8$. It was shown in \cite{tharaka_icc06}
that performance of UCA antenna configuration is less sensitive to
change of mean AOD $(\maod)$ and mean AOA $(\maoa)$. Therefore,
without loss of generality, we set $\maod=\maoa = 90^\circ$. Also,
in this simulation, we set transmitter angular spread
$\sigma_t=10^\circ$ and receiver angular spreads
$\sigma_r=\{30^\circ,10^\circ\}$. For comparison, also shown is the
average mutual information of the $3\times{3}$ i.i.d. MIMO channel.
We observe that when $\sigma_r=30^\circ$, both models give very
similar performance for all SNRs. When the angular spread at the
receiver is small, e.g. $\sigma_r=10^\circ$, we can observe that the
Kronecker model gives slightly higher performance than the
non-separable model for higher SNRs. However, the margin of capacity
over estimation is insignificant in comparison with the i.i.d.
channel capacity performance. \textit{Therefore, Kronecker model
provides a good estimation to the actual scattering channel when the
joint scattering distribution is uni-modal.} Reasoning for this
claim will be discussed in the next section.

\subsection{Capacity in Multi-Modal Bivariate Azimuth
Fields}\label{sec:performance_multimodal} 
\begin{figure}[h]
\centering
\includegraphics[width=\pictwidth]{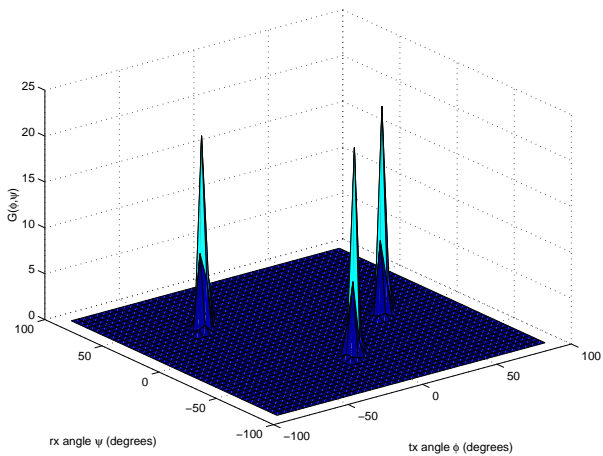}
\caption{An example multi-modal bivariate Gaussian distributed
azimuth field.\vspace{-0.5cm}} \label{fig:muli_modal_gaussian_apsd}
\end{figure}

A multi-modal azimuth
power distribution arises when there are two or more strong
multipaths exist in a fading channel. This may be the result of
multiple remote macroscopic scattering clusters, for instance. A
multi-modal bivariate distribution can be constructed from a mixture
of uni-modal bivariate distributions. Fig.
\ref{fig:muli_modal_gaussian_apsd} shows a multi-modal bivariate
Gaussian distributed azimuth field with 3 modes centered around
$(\maod,\maoa) =
\{(-40^\circ,40^\circ),(0^\circ,-40^\circ),(50^\circ,0^\circ)\}$,
each mode with angular spreads $\sigma_r=\sigma_t=5^\circ$ and
$\rho=0.8$. Note that, in this case the effective angular spreads at
the receiver and transmitter are larger than $5^\circ$.

\begin{figure}[h]
\centering
\includegraphics[width=\pictwidth]{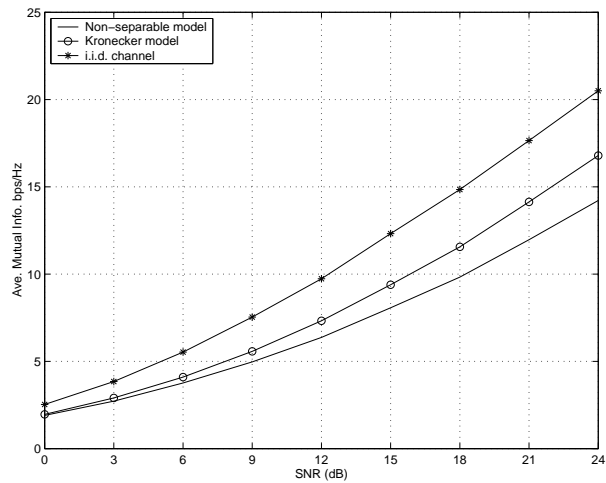}
\caption{Average mutual information of 3-transmit UCA and 3-receive
UCA MIMO system for separable and non-separable scattering channel
considered in Fig.
\ref{fig:muli_modal_gaussian_apsd}.\vspace{-0.5cm}}
\label{fig:muli_modal_gaussian_apsd_capacity}
\end{figure}

\begin{figure}[h]
\centering
\includegraphics[width=\pictwidth]{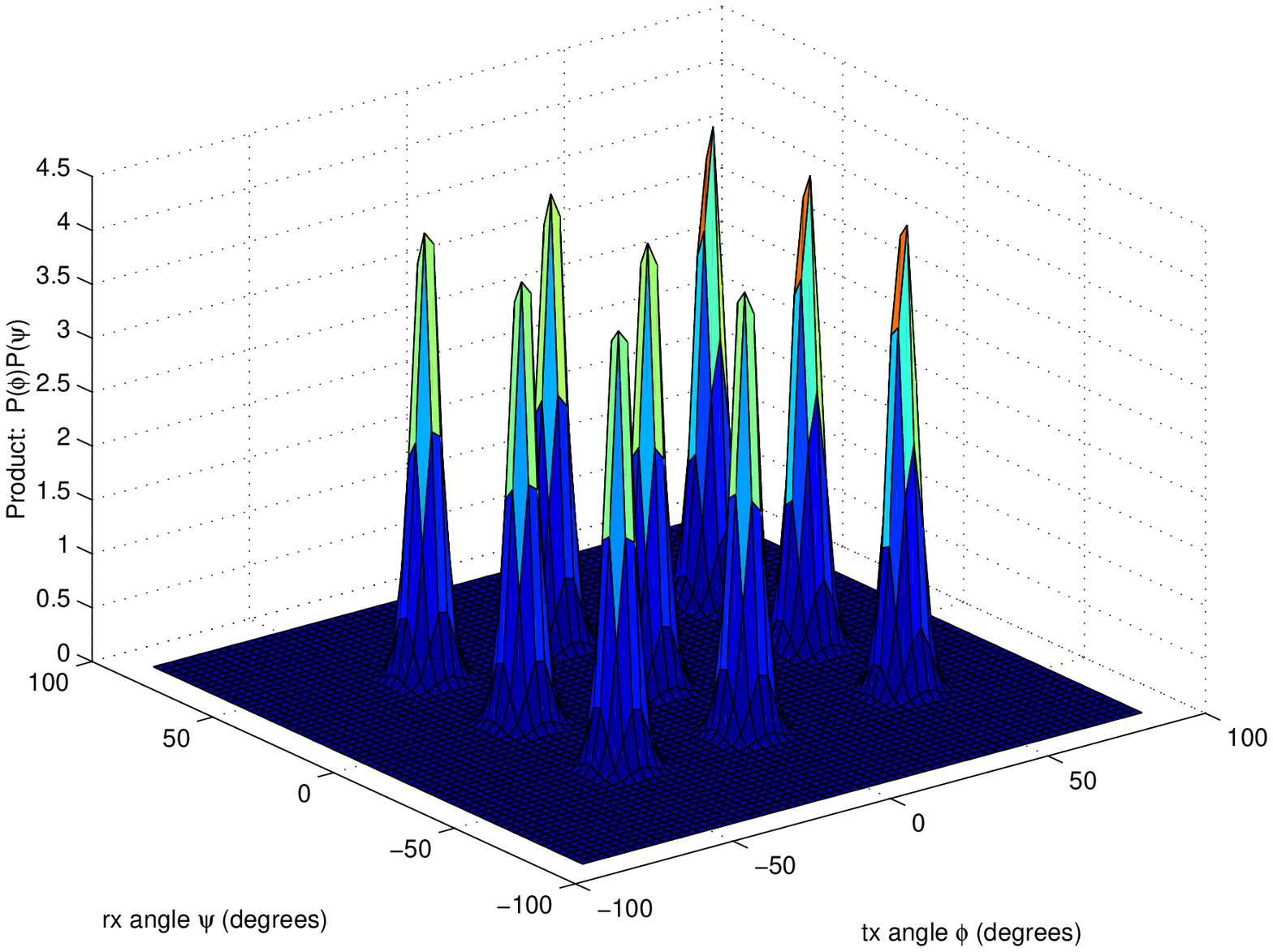}
\caption{Kronecker model PSD
$\widetilde{G}(\aod,\aoa)=\powd_{Tx}(\aod)\powd_{Rx}(\aoa)$ of the
non-separable scattering distribution considered in Fig.
\ref{fig:muli_modal_gaussian_apsd}.\vspace{-0.5cm}}
\label{fig:multi_modal_product_apsd}
\end{figure}

We now consider the $3{\times}3$ antenna configuration setup
discussed in the previous example. Fig.
\ref{fig:muli_modal_gaussian_apsd_capacity} shows the average mutual
information of it applied on the multi-modal scattering distribution
shown in Fig. \ref{fig:muli_modal_gaussian_apsd}. It is observed
that Kronecker model tends to overestimate the average mutual
information at high SNRs. Unlike in the uni-modal case considered
previously, the margin of error seen here is quite significant,
especially at high SNRs. Following the analysis given in
\cite{pollock_kronecker}, we now provide reasons for Kronecker model
to overestimate the mutual information for the scattering
distribution shown in Fig. \ref{fig:multi_modal_product_apsd}.

The joint PSD of the Kronecker model is given by
$\widetilde{G}(\aod,\aoa)=\powd_{Tx}(\aod)\powd_{Rx}(\aoa)$, where
$\powd_{Tx}(\aod)$ and $\powd_{Rx}(\aoa)$ are the transmit and
receive power distributions, generated by marginalizing
$G(\aod,\aoa)$. Fig. \ref{fig:multi_modal_product_apsd} shows the
Kronecker model PSD $\widetilde{G}(\aod,\aoa)$ of the scattering
channel considered in Fig. \ref{fig:muli_modal_gaussian_apsd}.
Comparing Fig. \ref{fig:multi_modal_product_apsd} with Fig.
\ref{fig:muli_modal_gaussian_apsd} we can observe that
$\widetilde{G}(\aod,\aoa)$ consist of six  extra modes,
corresponding to additional six scattering clusters. Therefore,
Kronecker model introduces virtual scattering clusters located at
the intersection of the actual scattering clusters. As a result,
Kronecker model will increase the effective angular spread at the
transmit and receive apertures (lower modal correlation) and hence
improved system performance. \textit{Therefore, the popular
Kronecker model does not model the MIMO channel accurately when
there exist multiple scattering clusters in the channel.} These
observations match the measurement results published in
\cite{bonek_kroneker_measurements}.

\begin{figure}[h]
\centering
\includegraphics[width=\pictwidth]{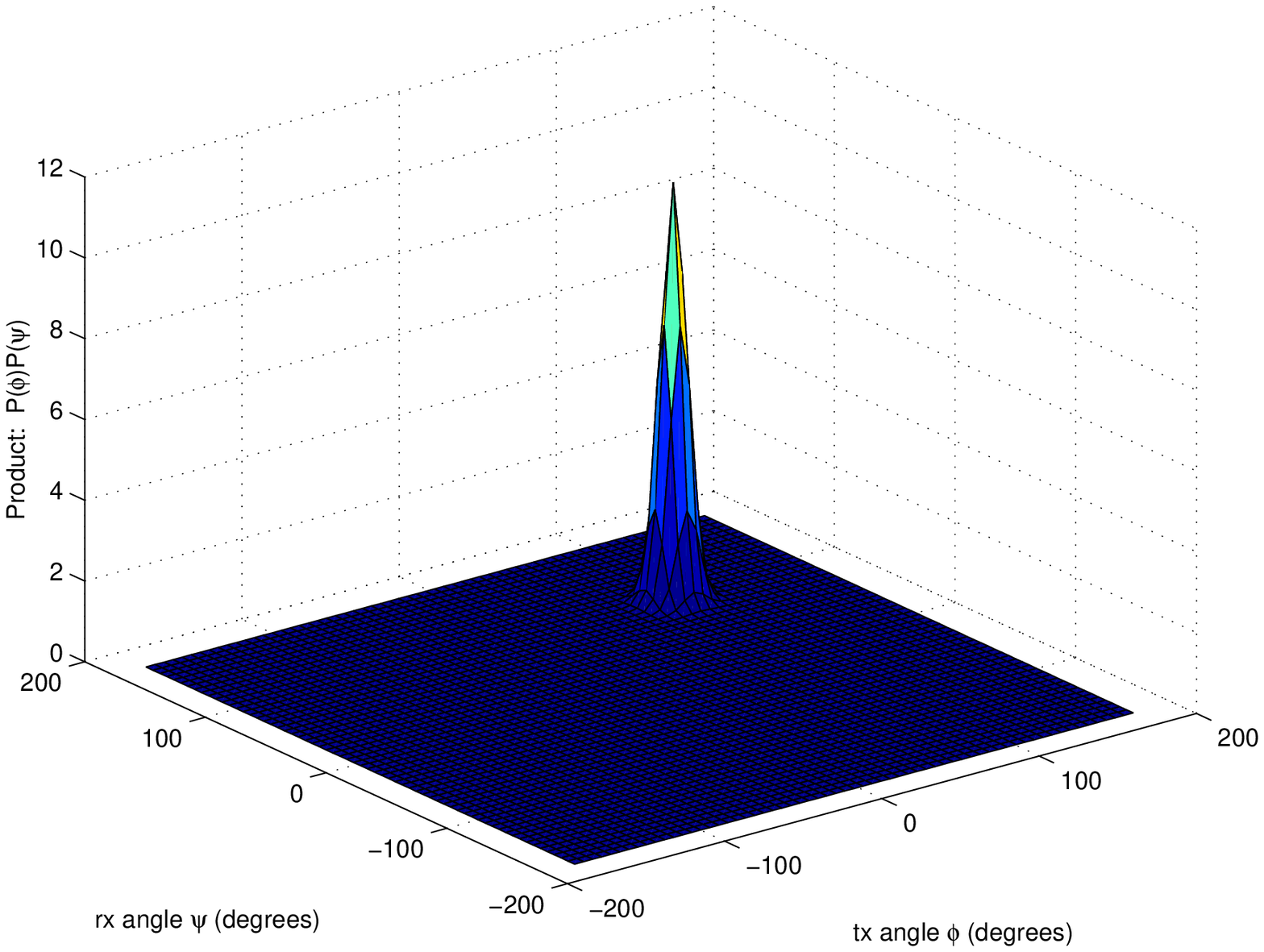}
\caption{Kronecker model PSD
$\widetilde{G}(\aod,\aoa)=\powd_{Tx}(\aod)\powd_{Rx}(\aoa)$ of the
uni-modal non-separable scattering distribution used in the first
example to obtain the results in Fig.
\ref{fig:unimodal_gauss_cap_sigr_30_10} for
$\sigma_r=10^\circ$.\vspace{-0.5cm}}
\label{fig:uni_modal_product_apsd}
\end{figure}

Now we consider the uni-modal PSD used in our first simulation
example. Fig. \ref{fig:uni_modal_product_apsd} shows the
corresponding Kronecker Model PSD $\widetilde{G}(\aod,\aoa)$ for
this channel, for $\sigma_r=10^\circ$ and $\sigma_t=10^\circ$. In
this case the Kronecker model does not introduce any additional
virtual scattering clusters into the channel. As a result, no
 increase in the number of multipaths of the channel,
hence both models give very similar performance.


\section{Conclusions}\label{sec:conclusion}
We presented a MIMO channel correlation model which is capable of
capturing antenna geometry and joint correlation properties of both
link ends. The scattering environment surrounding the transmitter
and receiver apertures is modeled using a bi-angular power
distribution. We use the covariance between transmit and receive
angles to control the joint correlation properties between transmit
and receive angular power distributions.

We showed that 2-D Fourier series coefficients of the bi-angular
power distribution and transmit and receive antenna sampling points
contribute to the entries of the correlation matrix. We proposed
several bi-angular power distributions and their 2-D Fourier series
coefficients in closed form.

Using the proposed model, we show that Kronecker model is a good
approximation to the actual channel when the scattering channel
consists of a single scattering cluster. In the presence of multiple
remote scattering clusters we show that Kronecker model over
estimates the performance of MIMO systems by introducing virtual
scattering clusters into the channel. Therefore, in this case,
Kronecker model cannot be used to represent the channel.


\end{document}